\documentstyle[12pt]{article}
\begin{document}

\title{Simply Modeling $\overline{B}\rightarrow \overline{K}^{\ast}
\gamma$}

\author{B. Holdom\thanks{holdom@utcc.utoronto.ca}
\hspace{0.5em}
and \hspace{0.4em}
M. Sutherland\thanks{marks@medb.physics.utoronto.ca}
     \and
     {\small {\it Department of Physics, University of Toronto}} \\
      {\small {\it 60 St. George St., Toronto, Ontario, Canada M5S 1A7}}
}
\date{\small {UTPT-93-22; hep-ph/9308313}}
\maketitle

\begin{abstract} A simple relativistic model of heavy-light
mesons is  applied to the rare decay $\overline{B}\rightarrow
\overline{K}^{\ast}\gamma$ in the standard model. We find
$\Gamma(\overline{B}\rightarrow
\overline{K}^{\ast}\gamma)/\Gamma(b\rightarrow s\gamma)= (17\pm4)$\%
and  BR$(\overline{B}\rightarrow
\overline{K}^{\ast}\gamma)=(4.8\pm1.9)\times 10^{-5}(|V_{cb}|/.04)^{2}$.
These numbers are reduced by only 20\% in the heavy-quark limit.
\end{abstract} \vspace{2ex}

We have recently developed a relativistic model of mesons containing a
heavy  quark $Q$ and a light antiquark $\overline{q}$
\cite{simply,hyperfine}.  Matrix elements for
$Q_{1}\overline{q}\rightarrow Q_{2}\overline{q}$ meson transitions are
represented by quark loop graphs with $\overline{Q}_{2}Q_{1}$-type
operator insertions on the heavy-quark line.  The external mesons are
joined to the loop with vertices of the form $Z^{2}/(-
k^{2}+\Lambda^{2})$, where $k$ is the light quark momentum.  These
vertices suppress large momentum flow into the light quark, which is  the
essential physical effect of the light quark wave function.  The  $Z$'s
and $\Lambda$'s are different for each meson flavor and spin,  but are
not arbitrary parameters.  They are fixed in terms of the  heavy and
light masses appearing in the standard quark propagators  by requiring
the meson self-energies to vanish and have unit slope at the
physically-measured meson masses.

Central to the application of a free quark model is the assumption that
QCD confinement is characterized by somewhat smaller momentum scales than
typical light quark momenta in a heavy-light meson.  This situation is
realized by our model and it thus suggests the possibility that
confinement may not play an essential role.  Indeed, our model results
are obtained by simply dropping the imaginary parts arising from the free
quark loop diagrams.  The success or failure of such a model will shed
light on the role played by confinement.  Thus far we have found that the
model yields a differential decay spectrum for  $\overline{B}\rightarrow
D^{\ast}\ell \overline{\nu}$ whose shape compares well with the data when
$m_{b}=4.80$ GeV, $m_{c}=1.44$  GeV and $m_{q}=250$ MeV \cite{linear}.
The model may also be expanded in inverse  powers of heavy-quark masses
and the vector- and axial-vector current form factors have been shown in
\cite{simply} to be consistent with all heavy-quark symmetry constraints
through order $1/m_{Q}$  \cite{Luke,NeuRieck}.

The purpose of this paper is to apply our model to the rare decay
$\overline{B}\rightarrow \overline{K}^{\ast}\gamma$ \cite{PJO} within the
context of the standard model.  We will be mainly interested in the
results of our full, unexpanded model, but we will also compare these
results to those obtained in the heavy-quark limit.  Because the strange
quark is not particularly heavy, there is no {\it a priori} reason to
expect there to be any resemblance.  We will see, however, that certain
quantities have  surprisingly small net corrections.

The relevant $\overline{s}b$-type operators are \cite{IL}
\begin{equation}
{\cal O}_{\mu}=\overline{s}i\sigma_{\mu\nu}q^{\nu}b \;\;\;{\rm  and}\;\;\;
 {\cal O}_{5\mu}= \overline{s} i\sigma_{\mu\nu}q^{\nu} \gamma_{5}  b .
\end{equation}
Form factors for  $\overline{B}(M,V)\rightarrow
\overline{K}^{\ast}(m,v,\varepsilon)$  may be defined by
\begin{equation}
\langle \overline{K}^{\ast}|{\cal O}_{\mu}|\overline{B} \rangle
=\sqrt{Mm}(M+m)h(\omega)
	\varepsilon_{\mu\nu\rho\sigma}\varepsilon^{\ast\nu}
		V^{\rho}v^{\sigma} \label{vector}
\end{equation}
\begin{eqnarray}
\langle \overline{K}^{\ast}|{\cal O}_{5\mu}|\overline{B} \rangle
 &=&  -i\sqrt{Mm} \left\{ \left[ (M-m)(\omega+1)g_{\mu\nu}
     -M(V+v)_{\mu} V_{\nu} \right] h_{5}(\omega) \right. \nonumber  \\
 & & \mbox{} + \left. \left[ (M+m)(\omega-1)g_{\mu\nu}
	+M(V-v)_{\mu} V_{\nu} \right] h_{5}^{\prime}(\omega)  \right\}
		\varepsilon^{\ast\nu} \label{axial}
\end{eqnarray}
where $\omega=V\cdot
v$ is the product of the mesons'  four-velocities.  The three
form factors $h$, $h_{5}$, and $h_{5}^{\prime}$ are not independent at the
physical recoil  point $\omega_{o}=(M^2+m^2)/2Mm =3.04$, where they satisfy
$h(\omega_{o})=h_{5}(\omega_{o})+(M-m)(M+m)^{-1}h_{5}^{\prime}
(\omega_{o})$.  In the heavy-quark limit, we find
$h(\omega)=h_{5}(\omega)=\xi(\omega)$ and  $h_{5}^{\prime}(\omega)=0$,
where $\xi$ is the same Isgur-Wise function appearing in meson
semileptonic decays.

As described below we will fix the $m_{b}$ and $m_{s}$ masses by going to
a point of minimal sensitivity  in the $m_{b}$-$m_{s}$ plane. It is
reassuring to find that this point occurs at the physically reasonable
values $m_{b}=4.83$ GeV and $m_{s}=.40$ GeV.  As in our earlier work, we
choose $m_{q}=250$ MeV as a reasonable light quark constituent mass
\cite{hyperfine}. This is somewhat smaller than the usual 330 MeV and it
models the fact that the actual momentum-dependent light quark mass has
fallen somewhat at typical light quark momenta in the loop.  We will find
little sensitivity to the choice of $m_q$. The vertices are given
explicitly by
\begin{equation}
\gamma_{5}\frac{Z_{B}^{2}}{-k^2+\Lambda_{B}^{2}} \;\; {\rm and} \;\;
-i\gamma_{\nu}\frac{Z_{K^{\ast}}^{2}}{-
k^{2}+\Lambda_{K^{\ast}}^{2}},
\label{df}
\end{equation}
and the various
constants are determined by the physical masses $M=5.279$ GeV and $m=.892$
GeV to be $\Lambda_{B}=.577$ GeV, $Z_{B}=1.013$ GeV,
$\Lambda_{K^{\ast}}=.617$ GeV and $Z_{K^{\ast}}=.871$ GeV.  These values
of $\Lambda$ characterize the typical light quark momenta.

The form factors computed using the full vertices of Eq. (\ref{df})   are
shown along with $\xi$ in Fig. 1.  Numerical results are shown in
(\ref{table1}) at the zero  recoil point $\omega=1$ and at the physical
recoil  point $\omega_{o}$.

\begin{equation} \begin{array}{ccccc} \omega & h & h_{5} & h_{5}^{\prime}
& \xi \\ \hline 1 & 1.27 & .964 & .207 & 1 \\ \omega_{o} & .262 & .205 &
.080 & .235  \end{array} \label{table1} \end{equation}

We see that the net deviations from the heavy-quark limit are in general
not  as large as one might have expected.   At the physical recoil point
$h$ and $h_{5}$ differ from $\xi$ by less than 15\%.  But this does not
mean that the $1/m_{s}$ expansion makes any sense; indeed if we write
\begin{equation} h(\omega_{o})=\xi(\omega_{o}) \left\{ 1+A/m_{s}+B/m_{b}
\right\}
\end{equation}
we find numerically that $A$ and $B$ are of order
$-150$ MeV.  This gives a correction going in the opposite direction from
the full result, and thus the higher order terms must be significant.

We also find that the leading order corrections to $h$ and $h_{5}$ vanish
at zero recoil in the model; this is analogous to Luke's theorem
\cite{Luke}.  In  particular, $h(1)-1=.27$ is entirely due to effects at
order  $1/m_{Q}^{2}$ and beyond.

The $b\rightarrow s\gamma$ vertex in the effective theory obtained  by
integrating out the $W$ boson and the top quark in the standard  model is
\cite{IL}
\begin{equation} \Gamma_{\mu} = \kappa [(1+r){\cal
O}_{\mu}+(1-r){\cal O}_{5\mu}] , \label{vertex}
\end{equation}
where
$\kappa=eG_{F} V_{ts}^{\ast}V_{tb}\overline{F}_{2}m_{b}/8\sqrt{2}\pi^{2}$
and $r=m_{s}/m_{b}$.  The coefficient $\overline{F}_{2}$ depends on the
mass $m_{t}$ of the top quark and contains the effects of QCD scaling
from $\mu=M_{W}$ down to $\mu=m_{b}$.  In the leading logarithmic
approximation it is given by \cite{GSW}
\begin{equation}
\overline{F}_{2}\approx \eta^{-16/23}\left\{
F_{2}(m_{t}^{2}/M_{W}^{2})+116(\eta^{10/23}-1)/135+
	58(\eta^{28/23}-1)/189 \right\}, \label{scaling}
\end{equation}
where $\eta=\alpha_{s}(m_{b})/\alpha_{s}(M_{W})$ and \cite{IL}
\begin{equation}
F_{2}(x)=\frac{8x^{3}+5x^{2}-7x}{12(x-1)^{3}}
	-\frac{3x^{3}-2x^{2}}{2(x-1)^{4}}\ln x . \label{F2}
\end{equation}
(We note that results in the next-to-leading-logarithmic approximation have
been computed and are not drastically different \cite{Misiak}.)

Using $\eta=\ln (M_{W}/\Lambda_{QCD})/ \ln(m_{b}/\Lambda_{QCD})=1.96$
with $\Lambda_{QCD}=250$ MeV, we find $\overline{F}_{2}/F_{2}$=1.92 for
$m_{t}=135$ GeV.  This  is a well-known large enhancement factor from
short-distance QCD  \cite{GSW,GOSN}.   We note in passing that attempts
have  been made to estimate the contribution of internal  $\psi-\gamma$
conversion to  $\overline{B} \rightarrow \overline{K}^{\ast}\gamma$ via
vector meson dominance \cite{internal}. However, the Wilson coefficient
of the relevant four-quark operator at $\mu=m_{b}$ is suppressed by
roughly a factor of three compared with its value at $\mu=M_{W}$
\cite{GOSN}.  We therefore neglect this contribution compared with the
QCD-enhanced short-distance one.

The width for $\overline{B}\rightarrow \overline{K}^{\ast}\gamma$  is
given by
\begin{equation} \Gamma(\overline{B}\rightarrow
\overline{K}^{\ast}\gamma) =M^{3}|\kappa|^{2}(16\pi
R)^{-1}(1-R^2)^{3}(1+R)^{2}  (1+r^{2})h(\omega_{o})^{2},
\label{mesonwidth}
\end{equation}
where $R=m/M$. A fundamental quantity
is the ratio of the exclusive $\overline{B}\rightarrow
\overline{K}^{\ast}\gamma$ decay width to that of the inclusive
$\overline{B}\rightarrow X_{s}\gamma$ decay. This may be taken to be
equal to the quark-level $b\rightarrow s\gamma$ width, given by
\begin{equation} \Gamma(b\rightarrow s\gamma)
=m_{b}^{3}|\kappa|^{2}(4\pi)^{-1}(1-r^2)^{3}(1+r^2), \label{quarkwidth}
\end{equation}
where $r=m_{s}/m_{b}$.  (The corrections
to this relation were shown in \cite{chayetal} to be of order
$1/m_{b}^{2}$ and may be neglected here.) The Kobayashi-Maskawa elements,
top quark mass and QCD scaling  effects cancel in the ratio, and we find
\begin{equation} \Gamma(\overline{B}\rightarrow
\overline{K}^{\ast}\gamma)/\Gamma(b\rightarrow  s\gamma)=17\%.
\end{equation}

Because $V_{ts}^{\ast}V_{tb}$ is not directly measured, it is convenient
to use unitarity and the smallness of $V_{ub}$ to write
$V_{ts}^{\ast}V_{tb}\approx -V_{cs}^{\ast}V_{cb}$ and express the
branching ratio in terms of $V_{cb}$.  With $|V_{cs}|=.974$,
$\tau_{B}=1.5\times 10^{-12}$ s and $m_{t}=135$ GeV, we find
\begin{equation} {\rm BR}(\overline{B}\rightarrow
\overline{K}^{\ast}\gamma)=4.8\times 10^{-5}(|V_{cb}|/.04)^{2}.
\end{equation}
The uncertainties due to a $\pm100$ MeV shift in
$\Lambda_{QCD}$ and a $\pm 25$ GeV shift in $m_{t}$ are $\pm10$\% each.
The rate for $\overline{B}\rightarrow \overline{K}^{\ast}\gamma$ is
found to decrease by 20\% if the form factor $h(\omega_{o})$ in
(\ref{mesonwidth}) is replaced by its value in the heavy-quark limit.

The CLEO collaboration \cite{CLEO} has recently reported a branching
ratio of $(4.5\pm1.5\pm0.9)\times 10^{-5}$.  We show in Fig. 2 how
$|V_{cb}|$ is constrained by the data as a function of $m_{t}$.

We now discuss the sensitivity of these results to the quark masses.
When $h(\omega_{o})$ is plotted as a function of $m_{b}$ and $m_{s}$ with
$m_{q}$ held fixed at 250 MeV, there is a saddle point at $m_{b}=4.83$
GeV and $m_{s}=.40$ GeV.  This point represents the point of minimal
sensitivity to the choice of quark masses and we adopt it as our standard
reference point. In the circular region of radius 40 MeV centred at this
point, we find that $h(\omega_{o})$ varies by less than 7.5\% from its
value at the saddle point.  We thus estimate a 15\% uncertainty in the
branching ratio due to the $b$ and $s$ quark masses.  The heightened
sensitivity to the quark masses which occurs farther away from the saddle
point has been discussed in  \cite{linear} in the context of
$\overline{B}\rightarrow D^{(\ast)}$ semileptonic decays.  There, it was
stressed that the sensitivity is the expected result of constraining the
meson masses to their physical values, and not a breakdown of the heavy-quark
expansion in the model.  We also find that varying the light quark mass $m_{q}$
by 40 MeV, with the $b$ and $s$ masses fixed,
changes the branching ratio by less than 5\%.
It will be possible in the future to more accurately determine the quark masses
appropriate to the model from heavy-meson and -baryon semileptonic decays.

In conclusion, we find in our model  $\Gamma(\overline{B}\rightarrow
\overline{K}^{\ast}\gamma)/\Gamma(b\rightarrow  s\gamma)=(17\pm4)$\% and
BR$(\overline{B}\rightarrow  \overline{K}^{\ast}\gamma)=(4.8\pm1.9)\times
10^{-5}(|V_{cb}|/.04)^{2}$.  These compare well with values of
$(20\pm6)$\% and $(6.8\pm2.4)\times 10^{-5}(|V_{cb}|/.035)^{2}$ obtained
in one recent QCD sum rule approach \cite{Ball}, and $(17\pm5)$\% and
$(4\pm1)\times 10^{-5}$ in another \cite{CDNP}.

\vspace{2ex}

\noindent {\bf Acknowledgment}
\vspace{1ex}

This research was supported in part by the Natural Sciences and
Engineering Research Council of Canada.

\newpage \noindent{\bf Figure Captions} \vspace{6ex}

\noindent {\bf Figure 1:} Model results for form factors
$h$, $h_{5}$ and $h_{5}^{\prime}$ and Isgur-Wise function $\xi$.
The physical recoil point  is $\omega=\omega_{o}$.
\vspace{4ex}

\noindent {\bf Figure 2:} Region in $|V_{cb}|$-$m_{t}$ plane
(between dotted lines)
allowed by data.  The solid line corresponds to the central value of
the branching ratio.  The figure is plotted for $m_{b}=4.83$
GeV and $m_{s}=400$ MeV.

\newpage
%\end{document}
%%%%
%%%%  To run this as two separate documents, cut here and
%%%%  comment \end \documentstyle and \begin statements back in.
%%%%
%\documentstyle[12pt]{article}
%\begin{document}
\begin{center}
% GNUPLOT: LaTeX picture
\setlength{\unitlength}{0.240900pt}
\ifx\plotpoint\undefined\newsavebox{\plotpoint}\fi
\sbox{\plotpoint}{\rule[-0.150pt]{0.300pt}{0.300pt}}%
\begin{picture}(1500,900)(0,0)
\font\gnuplot=cmr12 at 12pt
\gnuplot
\sbox{\plotpoint}{\rule[-0.150pt]{0.300pt}{0.300pt}}%
\put(176.0,113.0){\rule[-0.150pt]{4.818pt}{0.300pt}}
\put(154,113){\makebox(0,0)[r]{0}}
\put(1416.0,113.0){\rule[-0.150pt]{4.818pt}{0.300pt}}
\put(176.0,222.0){\rule[-0.150pt]{4.818pt}{0.300pt}}
\put(154,222){\makebox(0,0)[r]{0.2}}
\put(1416.0,222.0){\rule[-0.150pt]{4.818pt}{0.300pt}}
\put(176.0,331.0){\rule[-0.150pt]{4.818pt}{0.300pt}}
\put(154,331){\makebox(0,0)[r]{0.4}}
\put(1416.0,331.0){\rule[-0.150pt]{4.818pt}{0.300pt}}
\put(176.0,440.0){\rule[-0.150pt]{4.818pt}{0.300pt}}
\put(154,440){\makebox(0,0)[r]{0.6}}
\put(1416.0,440.0){\rule[-0.150pt]{4.818pt}{0.300pt}}
\put(176.0,550.0){\rule[-0.150pt]{4.818pt}{0.300pt}}
\put(154,550){\makebox(0,0)[r]{0.8}}
\put(1416.0,550.0){\rule[-0.150pt]{4.818pt}{0.300pt}}
\put(176.0,659.0){\rule[-0.150pt]{4.818pt}{0.300pt}}
\put(154,659){\makebox(0,0)[r]{1}}
\put(1416.0,659.0){\rule[-0.150pt]{4.818pt}{0.300pt}}
\put(176.0,768.0){\rule[-0.150pt]{4.818pt}{0.300pt}}
\put(154,768){\makebox(0,0)[r]{1.2}}
\put(1416.0,768.0){\rule[-0.150pt]{4.818pt}{0.300pt}}
\put(176.0,877.0){\rule[-0.150pt]{4.818pt}{0.300pt}}
\put(154,877){\makebox(0,0)[r]{1.4}}
\put(1416.0,877.0){\rule[-0.150pt]{4.818pt}{0.300pt}}
\put(176.0,113.0){\rule[-0.150pt]{0.300pt}{4.818pt}}
\put(176,68){\makebox(0,0){1}}
\put(176.0,857.0){\rule[-0.150pt]{0.300pt}{4.818pt}}
\put(428.0,113.0){\rule[-0.150pt]{0.300pt}{4.818pt}}
\put(428,68){\makebox(0,0){1.5}}
\put(428.0,857.0){\rule[-0.150pt]{0.300pt}{4.818pt}}
\put(680.0,113.0){\rule[-0.150pt]{0.300pt}{4.818pt}}
\put(680,68){\makebox(0,0){2}}
\put(680.0,857.0){\rule[-0.150pt]{0.300pt}{4.818pt}}
\put(932.0,113.0){\rule[-0.150pt]{0.300pt}{4.818pt}}
\put(932,68){\makebox(0,0){2.5}}
\put(932.0,857.0){\rule[-0.150pt]{0.300pt}{4.818pt}}
\put(1184.0,113.0){\rule[-0.150pt]{0.300pt}{4.818pt}}
\put(1184,68){\makebox(0,0){3}}
\put(1184.0,857.0){\rule[-0.150pt]{0.300pt}{4.818pt}}
\put(1436.0,113.0){\rule[-0.150pt]{0.300pt}{4.818pt}}
\put(1436,68){\makebox(0,0){3.5}}
\put(1436.0,857.0){\rule[-0.150pt]{0.300pt}{4.818pt}}
\put(176.0,113.0){\rule[-0.150pt]{303.534pt}{0.300pt}}
\put(1436.0,113.0){\rule[-0.150pt]{0.300pt}{184.048pt}}
\put(176.0,877.0){\rule[-0.150pt]{303.534pt}{0.300pt}}
\put(806,-22){\makebox(0,0){$\omega=v_{1}\cdot v_{2}$}}
\put(1320,299){\makebox(0,0)[l]{$h$}}
\put(1320,252){\makebox(0,0)[l]{$\xi$}}
\put(1320,204){\makebox(0,0)[l]{$h_{5}$}}
\put(1320,152){\makebox(0,0)[l]{$h_{5}^{\prime}$}}
\put(1204,31){\makebox(0,0)[l]{$\omega_{o}$}}
\put(176.0,113.0){\rule[-0.150pt]{0.300pt}{184.048pt}}
\multiput(1307.14,298.13)(-1.026,-0.497){84}{\rule{0.689pt}{0.120pt}}
\multiput(1308.57,298.38)(-86.570,-43.000){2}{\rule{0.344pt}{0.300pt}}
\put(1222,256){\vector(-2,-1){0}}
\multiput(1299.73,251.13)(-4.091,-0.488){20}{\rule{2.475pt}{0.117pt}}
\multiput(1304.86,251.38)(-82.863,-11.000){2}{\rule{1.238pt}{0.300pt}}
\put(1222,241){\vector(-4,-1){0}}
\multiput(1304.47,204.38)(-2.115,0.494){40}{\rule{1.332pt}{0.119pt}}
\multiput(1307.24,203.38)(-85.235,21.000){2}{\rule{0.666pt}{0.300pt}}
\put(1222,225){\vector(-4,1){0}}
\multiput(1287.77,152.39)(-9.396,0.469){8}{\rule{5.355pt}{0.113pt}}
\multiput(1298.89,151.38)(-76.885,5.000){2}{\rule{2.678pt}{0.300pt}}
\put(1222,157){\vector(-4,1){0}}
\put(1204,48){\vector(0,1){65}}
\sbox{\plotpoint}{\rule[-0.400pt]{0.800pt}{0.800pt}}%
\put(176,659){\usebox{\plotpoint}}
\multiput(176,659)(7.742,-9.755){7}{\usebox{\plotpoint}}
\multiput(226,596)(8.806,-8.806){6}{\usebox{\plotpoint}}
\multiput(277,545)(9.536,-8.010){5}{\usebox{\plotpoint}}
\multiput(327,503)(10.425,-6.812){10}{\usebox{\plotpoint}}
\multiput(428,437)(11.248,-5.345){9}{\usebox{\plotpoint}}
\multiput(529,389)(11.751,-4.124){13}{\usebox{\plotpoint}}
\multiput(680,336)(12.146,-2.747){20}{\usebox{\plotpoint}}
\multiput(932,279)(12.334,-1.723){22}{\usebox{\plotpoint}}
\put(1204,241){\usebox{\plotpoint}}
\sbox{\plotpoint}{\rule[-0.150pt]{0.300pt}{0.300pt}}%
\put(176,813){\usebox{\plotpoint}}
\multiput(176.37,810.10)(0.497,-1.042){98}{\rule{0.120pt}{0.699pt}}
\multiput(175.38,811.55)(50.000,-102.549){2}{\rule{0.300pt}{0.350pt}}
\multiput(226.37,706.83)(0.498,-0.746){100}{\rule{0.120pt}{0.522pt}}
\multiput(225.38,707.92)(51.000,-74.916){2}{\rule{0.300pt}{0.261pt}}
\multiput(277.37,631.17)(0.497,-0.610){98}{\rule{0.120pt}{0.441pt}}
\multiput(276.38,632.08)(50.000,-60.085){2}{\rule{0.300pt}{0.221pt}}
\multiput(327.00,571.13)(0.587,-0.499){170}{\rule{0.427pt}{0.120pt}}
\multiput(327.00,571.38)(100.113,-86.000){2}{\rule{0.214pt}{0.300pt}}
\multiput(428.00,485.13)(0.843,-0.498){118}{\rule{0.580pt}{0.120pt}}
\multiput(428.00,485.38)(99.796,-60.000){2}{\rule{0.290pt}{0.300pt}}
\multiput(529.00,425.13)(1.201,-0.498){124}{\rule{0.794pt}{0.120pt}}
\multiput(529.00,425.38)(149.352,-63.000){2}{\rule{0.397pt}{0.300pt}}
\multiput(680.00,362.13)(1.974,-0.498){126}{\rule{1.256pt}{0.120pt}}
\multiput(680.00,362.38)(249.393,-64.000){2}{\rule{0.628pt}{0.300pt}}
\multiput(932.00,298.13)(3.178,-0.497){84}{\rule{1.973pt}{0.120pt}}
\multiput(932.00,298.38)(267.906,-43.000){2}{\rule{0.986pt}{0.300pt}}
\put(176,639){\usebox{\plotpoint}}
\multiput(176.37,636.95)(0.497,-0.700){98}{\rule{0.120pt}{0.495pt}}
\multiput(175.38,637.97)(50.000,-68.973){2}{\rule{0.300pt}{0.248pt}}
\multiput(226.37,567.37)(0.498,-0.529){100}{\rule{0.120pt}{0.393pt}}
\multiput(225.38,568.19)(51.000,-53.185){2}{\rule{0.300pt}{0.196pt}}
\multiput(277.00,514.13)(0.595,-0.497){82}{\rule{0.432pt}{0.120pt}}
\multiput(277.00,514.38)(49.103,-42.000){2}{\rule{0.216pt}{0.300pt}}
\multiput(327.00,472.13)(0.778,-0.498){128}{\rule{0.541pt}{0.120pt}}
\multiput(327.00,472.38)(99.877,-65.000){2}{\rule{0.271pt}{0.300pt}}
\multiput(428.00,407.13)(1.101,-0.497){90}{\rule{0.734pt}{0.120pt}}
\multiput(428.00,407.38)(99.477,-46.000){2}{\rule{0.367pt}{0.300pt}}
\multiput(529.00,361.13)(1.546,-0.497){96}{\rule{0.999pt}{0.120pt}}
\multiput(529.00,361.38)(148.926,-49.000){2}{\rule{0.500pt}{0.300pt}}
\multiput(680.00,312.13)(2.386,-0.498){104}{\rule{1.501pt}{0.120pt}}
\multiput(680.00,312.38)(248.884,-53.000){2}{\rule{0.751pt}{0.300pt}}
\multiput(932.00,259.13)(3.911,-0.496){68}{\rule{2.406pt}{0.120pt}}
\multiput(932.00,259.38)(267.005,-35.000){2}{\rule{1.203pt}{0.300pt}}
\put(176,226){\usebox{\plotpoint}}
\multiput(176.00,225.14)(2.851,-0.485){16}{\rule{1.742pt}{0.117pt}}
\multiput(176.00,225.38)(46.385,-9.000){2}{\rule{0.871pt}{0.300pt}}
\multiput(226.00,216.14)(3.288,-0.482){14}{\rule{1.988pt}{0.116pt}}
\multiput(226.00,216.38)(46.875,-8.000){2}{\rule{0.994pt}{0.300pt}}
\multiput(277.00,208.14)(3.708,-0.480){12}{\rule{2.218pt}{0.116pt}}
\multiput(277.00,208.38)(45.397,-7.000){2}{\rule{1.109pt}{0.300pt}}
\multiput(327.00,201.13)(5.183,-0.486){18}{\rule{3.105pt}{0.117pt}}
\multiput(327.00,201.38)(94.555,-10.000){2}{\rule{1.553pt}{0.300pt}}
\multiput(428.00,191.14)(6.536,-0.482){14}{\rule{3.863pt}{0.116pt}}
\multiput(428.00,191.38)(92.983,-8.000){2}{\rule{1.931pt}{0.300pt}}
\multiput(529.00,183.14)(8.653,-0.485){16}{\rule{5.108pt}{0.117pt}}
\multiput(529.00,183.38)(140.397,-9.000){2}{\rule{2.554pt}{0.300pt}}
\multiput(680.00,174.13)(12.959,-0.486){18}{\rule{7.635pt}{0.117pt}}
\multiput(680.00,174.38)(236.153,-10.000){2}{\rule{3.818pt}{0.300pt}}
\multiput(932.00,164.14)(17.642,-0.482){14}{\rule{10.275pt}{0.116pt}}
\multiput(932.00,164.38)(250.674,-8.000){2}{\rule{5.138pt}{0.300pt}}
\end{picture}
\vspace{24ex}

{\bf Figure 1}
\end{center}

\newpage
\begin{center}
% GNUPLOT: LaTeX picture
\setlength{\unitlength}{0.240900pt}
\ifx\plotpoint\undefined\newsavebox{\plotpoint}\fi
\sbox{\plotpoint}{\rule[-0.150pt]{0.300pt}{0.300pt}}%
\begin{picture}(1500,900)(0,0)
\font\gnuplot=cmr12 at 12pt
\gnuplot
\sbox{\plotpoint}{\rule[-0.150pt]{0.300pt}{0.300pt}}%
\put(220.0,113.0){\rule[-0.150pt]{4.818pt}{0.300pt}}
\put(198,113){\makebox(0,0)[r]{0.02}}
\put(1416.0,113.0){\rule[-0.150pt]{4.818pt}{0.300pt}}
\put(220.0,209.0){\rule[-0.150pt]{4.818pt}{0.300pt}}
\put(198,209){\makebox(0,0)[r]{0.025}}
\put(1416.0,209.0){\rule[-0.150pt]{4.818pt}{0.300pt}}
\put(220.0,304.0){\rule[-0.150pt]{4.818pt}{0.300pt}}
\put(198,304){\makebox(0,0)[r]{0.03}}
\put(1416.0,304.0){\rule[-0.150pt]{4.818pt}{0.300pt}}
\put(220.0,400.0){\rule[-0.150pt]{4.818pt}{0.300pt}}
\put(198,400){\makebox(0,0)[r]{0.035}}
\put(1416.0,400.0){\rule[-0.150pt]{4.818pt}{0.300pt}}
\put(220.0,495.0){\rule[-0.150pt]{4.818pt}{0.300pt}}
\put(198,495){\makebox(0,0)[r]{0.04}}
\put(1416.0,495.0){\rule[-0.150pt]{4.818pt}{0.300pt}}
\put(220.0,591.0){\rule[-0.150pt]{4.818pt}{0.300pt}}
\put(198,591){\makebox(0,0)[r]{0.045}}
\put(1416.0,591.0){\rule[-0.150pt]{4.818pt}{0.300pt}}
\put(220.0,686.0){\rule[-0.150pt]{4.818pt}{0.300pt}}
\put(198,686){\makebox(0,0)[r]{0.05}}
\put(1416.0,686.0){\rule[-0.150pt]{4.818pt}{0.300pt}}
\put(220.0,781.0){\rule[-0.150pt]{4.818pt}{0.300pt}}
\put(198,781){\makebox(0,0)[r]{0.055}}
\put(1416.0,781.0){\rule[-0.150pt]{4.818pt}{0.300pt}}
\put(220.0,877.0){\rule[-0.150pt]{4.818pt}{0.300pt}}
\put(198,877){\makebox(0,0)[r]{0.06}}
\put(1416.0,877.0){\rule[-0.150pt]{4.818pt}{0.300pt}}
\put(331.0,113.0){\rule[-0.150pt]{0.300pt}{4.818pt}}
\put(331,68){\makebox(0,0){100}}
\put(331.0,857.0){\rule[-0.150pt]{0.300pt}{4.818pt}}
\put(552.0,113.0){\rule[-0.150pt]{0.300pt}{4.818pt}}
\put(552,68){\makebox(0,0){120}}
\put(552.0,857.0){\rule[-0.150pt]{0.300pt}{4.818pt}}
\put(773.0,113.0){\rule[-0.150pt]{0.300pt}{4.818pt}}
\put(773,68){\makebox(0,0){140}}
\put(773.0,857.0){\rule[-0.150pt]{0.300pt}{4.818pt}}
\put(994.0,113.0){\rule[-0.150pt]{0.300pt}{4.818pt}}
\put(994,68){\makebox(0,0){160}}
\put(994.0,857.0){\rule[-0.150pt]{0.300pt}{4.818pt}}
\put(1215.0,113.0){\rule[-0.150pt]{0.300pt}{4.818pt}}
\put(1215,68){\makebox(0,0){180}}
\put(1215.0,857.0){\rule[-0.150pt]{0.300pt}{4.818pt}}
\put(1436.0,113.0){\rule[-0.150pt]{0.300pt}{4.818pt}}
\put(1436,68){\makebox(0,0){200}}
\put(1436.0,857.0){\rule[-0.150pt]{0.300pt}{4.818pt}}
\put(220.0,113.0){\rule[-0.150pt]{292.934pt}{0.300pt}}
\put(1436.0,113.0){\rule[-0.150pt]{0.300pt}{184.048pt}}
\put(220.0,877.0){\rule[-0.150pt]{292.934pt}{0.300pt}}
\put(23,495){\makebox(0,0){$|V_{cb}|$}}
\put(828,-22){\makebox(0,0){top quark mass (GeV)}}
\put(220.0,113.0){\rule[-0.150pt]{0.300pt}{184.048pt}}
\put(220,548){\usebox{\plotpoint}}
\multiput(220.00,547.13)(2.669,-0.494){40}{\rule{1.661pt}{0.119pt}}
\multiput(220.00,547.38)(107.553,-21.000){2}{\rule{0.830pt}{0.300pt}}
\multiput(331.00,526.13)(2.928,-0.493){36}{\rule{1.812pt}{0.119pt}}
\multiput(331.00,526.38)(106.239,-19.000){2}{\rule{0.906pt}{0.300pt}}
\multiput(441.00,507.13)(3.308,-0.492){32}{\rule{2.034pt}{0.119pt}}
\multiput(441.00,507.38)(106.779,-17.000){2}{\rule{1.017pt}{0.300pt}}
\multiput(552.00,490.13)(3.724,-0.491){28}{\rule{2.275pt}{0.118pt}}
\multiput(552.00,490.38)(105.278,-15.000){2}{\rule{1.138pt}{0.300pt}}
\multiput(662.00,475.13)(4.350,-0.490){24}{\rule{2.637pt}{0.118pt}}
\multiput(662.00,475.38)(105.528,-13.000){2}{\rule{1.318pt}{0.300pt}}
\multiput(773.00,462.13)(4.680,-0.489){22}{\rule{2.825pt}{0.118pt}}
\multiput(773.00,462.38)(104.137,-12.000){2}{\rule{1.413pt}{0.300pt}}
\multiput(883.00,450.14)(6.355,-0.485){16}{\rule{3.775pt}{0.117pt}}
\multiput(883.00,450.38)(103.165,-9.000){2}{\rule{1.888pt}{0.300pt}}
\multiput(994.00,441.13)(5.646,-0.486){18}{\rule{3.375pt}{0.117pt}}
\multiput(994.00,441.38)(102.995,-10.000){2}{\rule{1.688pt}{0.300pt}}
\multiput(1104.00,431.13)(5.698,-0.486){18}{\rule{3.405pt}{0.117pt}}
\multiput(1104.00,431.38)(103.933,-10.000){2}{\rule{1.703pt}{0.300pt}}
\multiput(1215.00,421.14)(8.193,-0.480){12}{\rule{4.789pt}{0.116pt}}
\multiput(1215.00,421.38)(100.060,-7.000){2}{\rule{2.395pt}{0.300pt}}
\multiput(1325.00,414.14)(7.185,-0.482){14}{\rule{4.238pt}{0.116pt}}
\multiput(1325.00,414.38)(102.205,-8.000){2}{\rule{2.119pt}{0.300pt}}
\sbox{\plotpoint}{\rule[-0.400pt]{0.800pt}{0.800pt}}%
\put(220,375){\usebox{\plotpoint}}
\multiput(220,375)(12.326,-1.777){10}{\usebox{\plotpoint}}
\multiput(331,359)(12.339,-1.683){8}{\usebox{\plotpoint}}
\multiput(441,344)(12.369,-1.449){9}{\usebox{\plotpoint}}
\multiput(552,331)(12.380,-1.351){9}{\usebox{\plotpoint}}
\multiput(662,319)(12.393,-1.228){9}{\usebox{\plotpoint}}
\multiput(773,308)(12.420,-0.903){9}{\usebox{\plotpoint}}
\multiput(883,300)(12.413,-1.006){9}{\usebox{\plotpoint}}
\multiput(994,291)(12.420,-0.903){9}{\usebox{\plotpoint}}
\multiput(1104,283)(12.435,-0.672){9}{\usebox{\plotpoint}}
\multiput(1215,277)(12.428,-0.791){9}{\usebox{\plotpoint}}
\multiput(1325,270)(12.435,-0.672){9}{\usebox{\plotpoint}}
\put(1436,264){\usebox{\plotpoint}}
\put(220,690){\usebox{\plotpoint}}
\multiput(220,690)(12.149,-2.736){10}{\usebox{\plotpoint}}
\multiput(331,665)(12.232,-2.335){9}{\usebox{\plotpoint}}
\multiput(441,644)(12.236,-2.315){9}{\usebox{\plotpoint}}
\multiput(552,623)(12.307,-1.902){9}{\usebox{\plotpoint}}
\multiput(662,606)(12.341,-1.668){9}{\usebox{\plotpoint}}
\multiput(773,591)(12.354,-1.572){9}{\usebox{\plotpoint}}
\multiput(883,577)(12.369,-1.449){9}{\usebox{\plotpoint}}
\multiput(994,564)(12.380,-1.351){8}{\usebox{\plotpoint}}
\multiput(1104,552)(12.413,-1.006){9}{\usebox{\plotpoint}}
\multiput(1215,543)(12.402,-1.127){9}{\usebox{\plotpoint}}
\multiput(1325,533)(12.429,-0.784){9}{\usebox{\plotpoint}}
\put(1436,526){\usebox{\plotpoint}}
\end{picture}
\vspace{24ex}

{\bf Figure 2}
\end{center}
\end{document}